\documentstyle[emulateapj]{article}

\tighten
\newcommand{\etal}{{\rm et al.~}} 
\newcommand{\Mpc}{$h^{-1}$~{\rm Mpc}}
\newcommand{\hmpc}{$h$~{\rm Mpc$^{-1}$}}

\def\apj{ApJ\ } 
\def\apjl{ApJL\ } 
\def\apjs{ApJS\ } 

\def\mn{MNRAS }

\begin{document}

\title{Optical and X-ray clusters as tracers of the supercluster-void
network. II The spatial correlation function}

\author{E. Tago\altaffilmark{1},
J. Einasto \altaffilmark{1},
M. Einasto\altaffilmark{1},
V. M\"uller\altaffilmark{2} \&  
H. Andernach\altaffilmark{3}}
\altaffiltext{1}{Tartu Observatory, EE-61602 T\~oravere, Estonia}
\altaffiltext{2}{Astrophysical Institute Potsdam, An der Sternwarte 16,
       D-14482 Potsdam, Germany}
\altaffiltext{3} {Depto.  de Astronom\'\i a, Univ.\ Guanajuato,
            Apdo.\ Postal 144, Guanajuato, C.P.\ 36000, GTO, Mexico}

\begin{abstract}

We study the space distribution of Abell and X-ray selected clusters
of galaxies from the ROSAT Bright Source Catalog, and determine
correlation functions for both cluster samples.  On small scales the
correlation functions depend on the cluster environment: clusters in
rich superclusters have a larger correlation length and amplitude than
all clusters.  On large scales correlation functions depend on the
distribution of superclusters.  On these scales correlation functions
for both X-ray and Abell clusters are oscillating with a period of
$\sim 115$~\Mpc.  This property shows the presence of a dominating
scale in the distribution of rich superclusters.

\end{abstract}

\section{Introduction}

The distribution of matter can be characterized by the power spectrum,
the correlation function, the void probability function, and other
suitable distribution functions.  On scales up to about $100 -
200$~\Mpc\ (in this paper we denote Hubble constant as $H_0 =
100~h$~km~s$^{-1}$~Mpc$^{-1}$) the power spectrum of galaxies and
clusters of galaxies has been determined in a number of studies using
available surveys; for recent reviews about the power spectrum see
Peacock \& Dodds (1994), Vogeley (1998), Einasto \etal (1999a,
hereafter E99a), Retzlaff (1999), Schuecker \etal 2000, and Miller and
Batuski (2000).  These studies show that on smaller scales the power
spectrum exhibits an almost exact power law of index $n \simeq -1.9$,
has a maximum at $k \simeq 0.05 \pm 0.01$~\hmpc, or wavelength
$\lambda \approx 120$~\Mpc, and probably approaches the
Harrison-Zeldovich spectrum with power index $n=1$ on very large
scales.  The exact shape of the power spectrum around the maximum and
beyond is known with rather low accuracy.  Einasto \etal (1997a) and
Retzlaff \etal (1998) find that the power spectrum of Abell clusters
has a sharp peak near $k=0.05$, whereas Miller \& Batuski (2000) did
not detect a strong feature on this scale.

Another popular method to characterize the large-scale structure is
the use of the correlation function; for recent studies, including on
X-ray sources, see Romer \etal (1994), Einasto \etal (1997b, hereafter
E97b), Guzzo (1999), Abadi \etal (1998), Lee \& Park (1999), Guzzo
\etal (1999), Tesch \etal (2000), Moscardini \etal (2000a,b) and Collins
\etal (2000). In most studies the correlation function has been
investigated only in a relatively small distance range up to about
100~\Mpc.  In these studies a log-log representation was used, thus it
was possible to investigate the function only in the range of
separations where the correlation function is positive.  On larger
scales, using a non-logarithmic representation, E97b and Broadhurst \&
Jaffe (1999) found oscillations of the spatial correlation
function.  Oscillations are manifested by alternating peaks and valleys
of the functions with a period of $\approx 120$~\Mpc.

\begin{figure*}[ht]
\vspace*{10.0cm} 
\figcaption{Sky coverage presented in 
Galactic coordinates for three X-ray selected clusters samples: + for
Schwope \etal (RBS), x for Ebeling \etal (RASS BCS), and triangles for
De Grandi \etal (RASS1 Bright Sample).  Asterisks appear when clusters
from the two first samples coincide.  The dashed line denotes the 
celestial equator (declination $\delta = 0^{\circ}$).   }  
\includegraphics{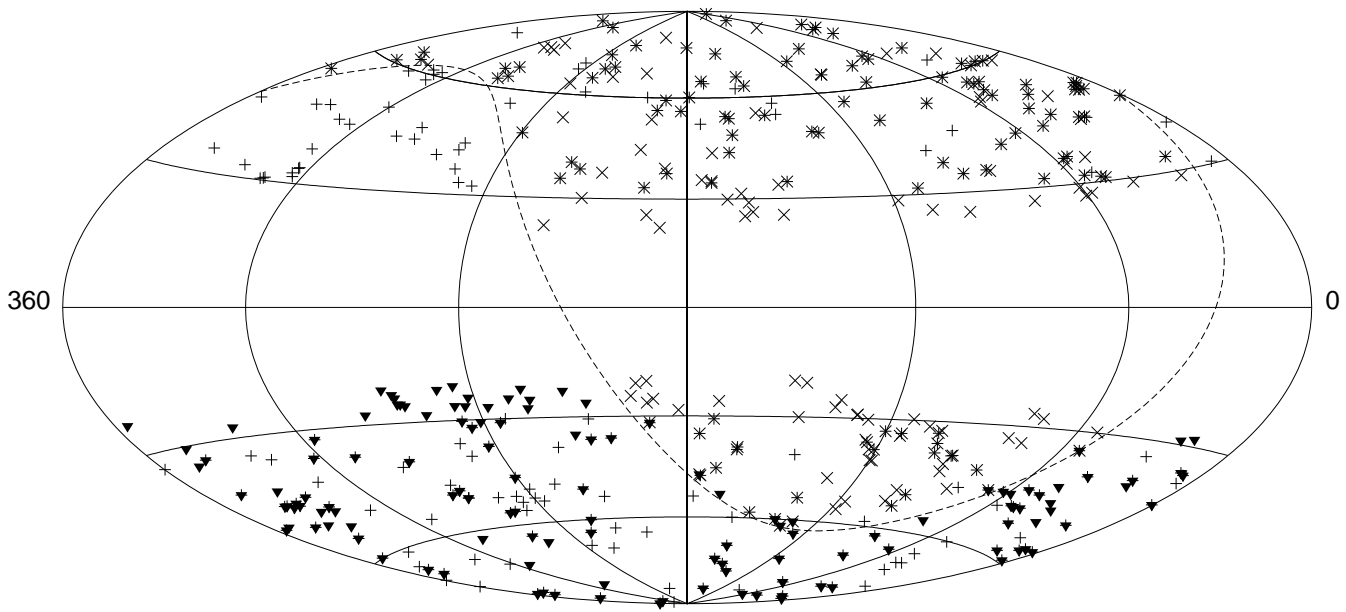}
\label{figure1}
\end{figure*}

The primary goal of the present paper is to compare the distribution
of optically selected Abell (1958), and Abell, Corwin \& Olowin (1989)
clusters with the distribution of X-ray selected clusters.  Of
particular interest is the comparison of distributions on large scales
and to check the presence of a preferred scale around 120~\Mpc.  We
shall use the correlation function analysis as this function is more
sensitive to the presence of the regularity in the distribution of
clusters (E97b). 

The ROSAT all-sky X-ray imaging survey has provided a unique
possibility to compile samples of clusters independently of the
previous optical catalogs.  There exist several cluster surveys based
on ROSAT observations: the ROSAT-ESO Flux Limited X-ray (REFLEX)
cluster survey (Guzzo \etal 1999, Collins \etal 2000, Schuecker \etal
2000), the ROSAT Brightest Cluster Sample (BCS, Ebeling \etal 1998,
2000), the RASS1 Bright Sample southern survey (De Grandi \etal
1999), and the ROSAT Bright Survey (RBS, Schwope \etal 2000); the
latter is based on the ROSAT all-sky Bright Source Catalog (RBSC,
Voges \etal 1999).  We shall use the sample of X-ray clusters by
Schwope \etal (2000) since this is the only available all-sky survey
of X-ray selected clusters so far. Redshifts are presently available
for almost all objects in this catalog.  We use a new version of the
compilation of redshifts of Abell clusters to study the distribution
of X-ray selected clusters in superclusters determined by Abell
clusters (Einasto \etal 2001a, Paper I).

The paper is organized as follows. In the next Section we shall
describe X-ray and optically selected samples of clusters of galaxies
used in the study.  In Section 3 we shall determine relevant selection
functions and calculate the correlation functions for X-ray and Abell
clusters.  Here we determine the correlation function for all clusters
of Abell and X-ray selected samples, as well as correlation functions
of clusters located in superclusters.  Lists of clusters in
superclusters were published in Paper I.  In Section 4 we present the
analysis of the nearest neighbor distribution for X-ray cluster systems,
and in Section 5 we briefly discuss our results and give conclusions.

\begin{table*}
\begin{center}
\caption[dummy]{\centerline {Properties of the samples of X-ray clusters}}
\label{tab:param1}
\begin{tabular}{lclcclllll}
\\ \hline \hline \\ Sample & $N_{total}$ & Sky area & Energy band &
 $CR_{limit}$ & Flux limit &  $r_0$ & $\gamma$ 

\\
& & & keV & cts/s & 
 $10^{-12}$ergs~cm$^{-2}$s$^{-1}$ & $h^{-1}$ Mpc & 
 \\ \hline \\ 
Ebeling 1998 & 201 & $|b_g| > 20^{\circ}, \delta > 0^{\circ}$ &
$0.1-2.4$ & & 4.4 & 33.0 & 1.82 \\ 
De Grandi 1999 & 130 & $|b_g| > 20^{\circ}, \delta < 2.5^{\circ}$ & 
$0.5-2.0$ & & $3-4$ &  21.5 & 2.11 \\ 
Schwope 2000 & 302 & $|b_g| > 30^{\circ}$ & $0.5-2.0$ & 0.2 & 2.4 & 25 & 2.2 
\\ \\ \hline
\end{tabular}
\end{center}
\tablenotetext { } {\hskip 0.3truecm Note: for the correlation
function parameters of the first two samples see Moscardini \etal
2000b and references therein}
\end{table*}

\section{Data}

\subsection{X-ray selected samples}

\begin{figure*}[ht]
\vspace*{7.0cm} \figcaption{Left panel: the X-ray luminosity
vs. distance for RBS clusters.  Right panel: The number of X-ray
clusters in superclusters of different multiplicity $k$; isolated
clusters are shown for comparison. System with the largest number of
X-ray clusters is the Hercules supercluster }
\includegraphics{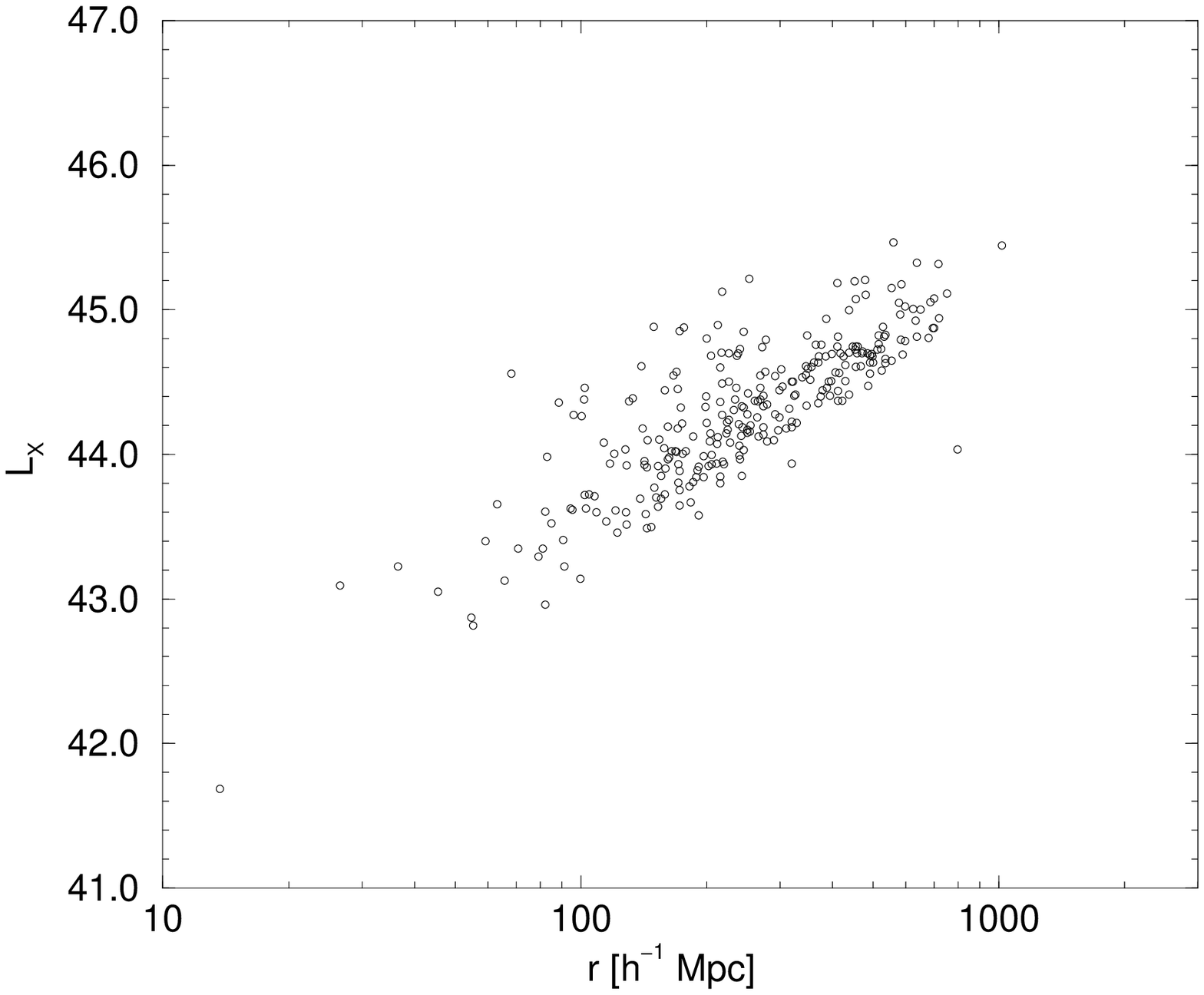} 
\includegraphics{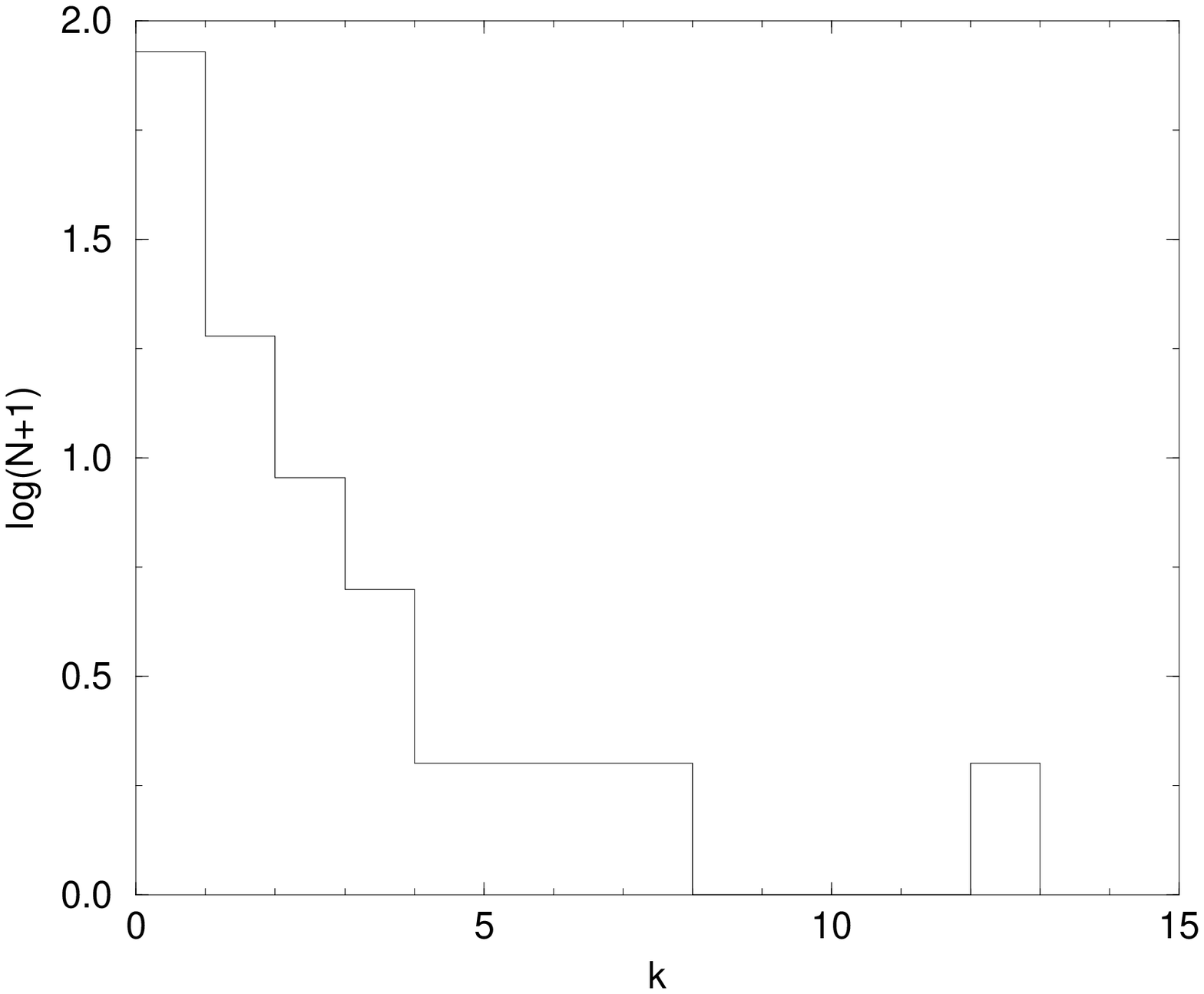}
\label{figure2}
\end{figure*}

ROSAT observations were made with the Position Sensitive Proportional
Counter in the broad ($0.1 - 2.4$~keV), soft ($0.1 - 0.4$~keV) and
hard energy band ($0.5 - 2.0$~keV) (Voges \etal 1999). The ROSAT
All-Sky Survey Bright Source Catalog (RASS-BSC) contains in total
18,811 sources down to a limiting count rate 0.05 cts/s.  Presently
three surveys of X-ray selected clusters of galaxies based on ROSAT
observations are available, those by Ebeling \etal (1998, 2000), De
Grandi \etal (1999), and Schwope \etal (2000). In this paper we use
the ROSAT Bright Survey (RBS, Schwope \etal 2000). This catalog of all
high-Galactic latitude RASS sources with PSPC count-rate above
0.2\,s$^{_1}$ contains a total of 302 X-ray clusters and is the only
available survey covering the whole sky (excluding the zone of
avoidance).  We denote this cluster sample as the ``RBS'' sample.

Figure~1 shows the sky distribution of clusters for these samples.
Here we can compare the sample selection criteria and the sky coverage
for these catalog to estimate the best sample for our purposes.
Table~1 presents some parameters for these samples.  We see that the
RBS sample is the only one that covers both hemispheres, although a
wider zone of avoidance has been used -- an advantage to reduce latitude
selection effects.

We have used a restricted sub-sample of the RBS catalog.  Figure~2
(left-hand panel) presents the X-ray luminosity versus distance
distribution for the RBS cluster sample. As in Paper I all distances
have been calculated using the formula by Mattig (1958). The sharp
edge at the lower part of the distribution corresponds to the survey
flux (apparent X-ray magnitude) limit.  For the calculation of the
correlation function and multiplicity function we have restricted our
sample to limit of X-ray {\em absolute luminosity} $ L_x \geq
10^{43}$~erg/s.  Up to the distance where this luminosity equals the
flux limit used (approximately 100~\Mpc), our sample is volume
limited. At larger distances the sample is flux limited, and thus
subject to a bias similar to the Malmquist bias in optically selected
galaxy catalogs.  Due to this effect the sample becomes very diluted
on large scales.  To decrease the influence of this effect we have cut
the sample at a distance of 250~\Mpc. Our previous experience has
shown that very diluted samples do not represent well the distribution
of high-density regions in the universe (E97b).  The mentioned
restriction in luminosity and distance provided a more homogeneous
sample of 137 X-ray clusters which we shall refer to as the
``RBS.250'' sample.

\subsection{Abell cluster samples}

For comparison with X-ray clusters we shall study the clustering
properties of optically selected Abell clusters of galaxies (Abell
1958, Abell, Corwin \& Olowin 1989).  Here we use the March 1999
version of the redshift compilation for Abell clusters described by
Andernach \& Tago (1998).  This sample was described in Paper I. It
contains all clusters of richness class $R \geq 0$ with measured or
estimated redshifts not exceeding $z_{lim}=0.13$; beyond this limit
the fraction of clusters with measured redshifts becomes small.  The
sample contains 1662 clusters, 1071 of which have measured redshifts.
We denote this sample of Abell clusters ACO.A1; A for all and 1 for
lower limit of supercluster richness of clusters included into the
sample.  Einasto \etal (2001b, Paper III) present the correlation
functions of Abell clusters for full samples (including clusters with
estimated redshifts) and for samples which include only clusters with
measured redshifts.

\subsection{Superclusters of Abell and X-ray clusters}

In our previous papers (e.g. E97d) we have shown that on large scales
the correlation function characterizes the distribution of
high-density regions.  High-density regions are determined essentially
by clusters of galaxies located in superclusters. We defined
high-density regions, applying cluster analysis, as superclusters of
Abell clusters, and as superclusters of X-ray clusters.

\begin{table*}
\begin{center}
\caption{\centerline{Parameters of the selection function and the 
correlation function for various samples}}
\label{tab:param2}
\begin{tabular}{lcccrccc}\\ 
\hline \hline \\ Sample & $\sin b_0$ & $d_0$ & $p$ & $N$ & $r_0$ &
$\gamma$ & $P$ \cr 
& & $h^{-1}$ Mpc & & & $h^{-1}$ Mpc & & $h^{-1}$ Mpc 
\\ \hline \\
RBS.250 & 0.22 & 100 & 1.0 & 137  & 24.7 $\pm 1.9$  & 2.2 $\pm$ 0.3 & 115 
$\pm$ 7 \cr 
RBS.XSC & 0.25 & 100 & 1.5 & 91   & 36.8 $\pm 2.1$ & 2.4 $\pm$ 0.2 & 109 
$\pm$ 19 \cr 
RBS.A8& 0.25 & 100 & 1.5 & 56   & 49.9 $\pm 6.3$ & 2.2 $\pm$ 0.3 & 111
 $\pm$ 18 \cr 
ACO.A1   & 0.25 & 100 & 1.2 & 1662 & 17.6 $\pm 1.2$ & 1.7 $\pm$ 0.2 & 113 
$\pm$ 28 \cr 
ACO.A8  & 0.40 & 100 & 1.0 & 373  & 36.7 $\pm 5.1$ & 2.3 $\pm$ 0.4 & 115 
$\pm$ 17 \cr \\ 
\hline
\end{tabular}
\tablenotetext { } {\hskip 1truecm Note: in the last three columns
1~$\sigma$ errors are presented}
\end{center}
\end{table*}

We searched for superclusters on the basis of both RBS.250 and Abell
cluster samples using the friend-of-friend algorithm with a
neighborhood radius of $24$~\Mpc; for details see Einasto \etal
(1994, hereafter EETDA), Einasto \etal (1997d, hereafter E97d), Paper
I and references therein. This radius was chosen according to the
clustering properties of Abell clusters: at this radius the
friend-of-friend algorithm connects clusters into systems so that we
obtain superclusters as the largest still relatively isolated systems.
We applied the same neighborhood radius to the X-ray cluster sample
and search for superclusters, i.e.\ systems of clusters with at least
two members.  A supercluster is called ``rich'' if it contains at
least 4 member clusters, and ``very rich'' if it has at least 8 member
clusters (E97d).  The sample of Abell clusters which belong to the
very rich superclusters is denoted as ACO.A8 (see Paper I for details
and the list of superclusters based on Abell clusters).

\begin{figure*}[ht]
\vspace*{7.0cm} 
\figcaption{Selection functions for X-ray and ACO
clusters in distance (left-hand panel) and in Galactic latitude
(right-hand panel). Dotted lines show selection function for the
sample ACO.A1.  Solid lines show the distribution of clusters in the whole
sample RBS.  Dashed lines show distribution of clusters in RBS.250
selected by limiting X-ray luminosity ($\log L_x > 43$, in both panels)
and distance ($R < 250$~\Mpc, in the right-hand panel). In the right-hand
panel the density is given in units of density at the Galactic
pole (except for the RBS.250 which is scaled to the RBS
density). Linear approximations are given for the  respective latitude
selection functions. }
\includegraphics{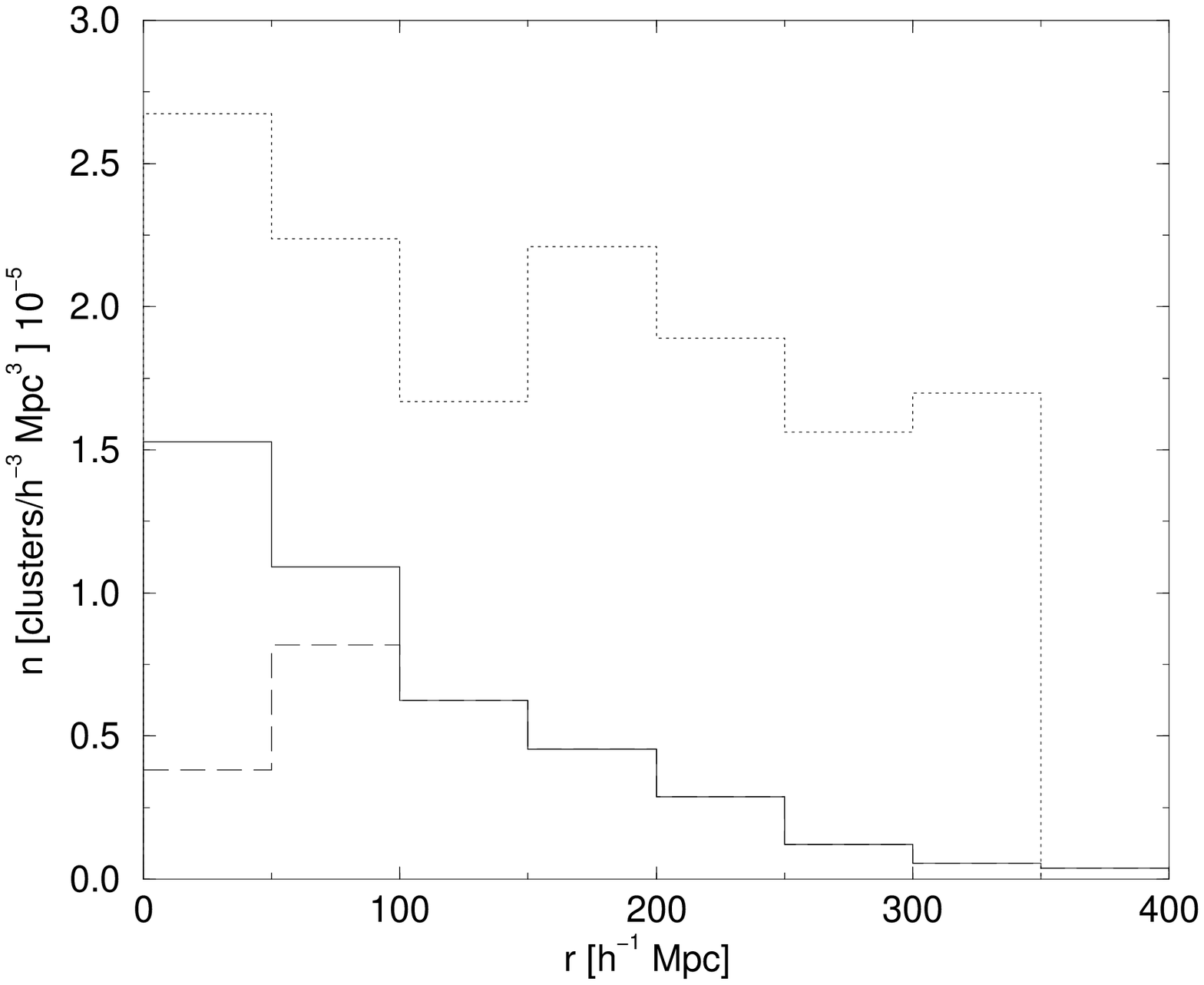}
\includegraphics{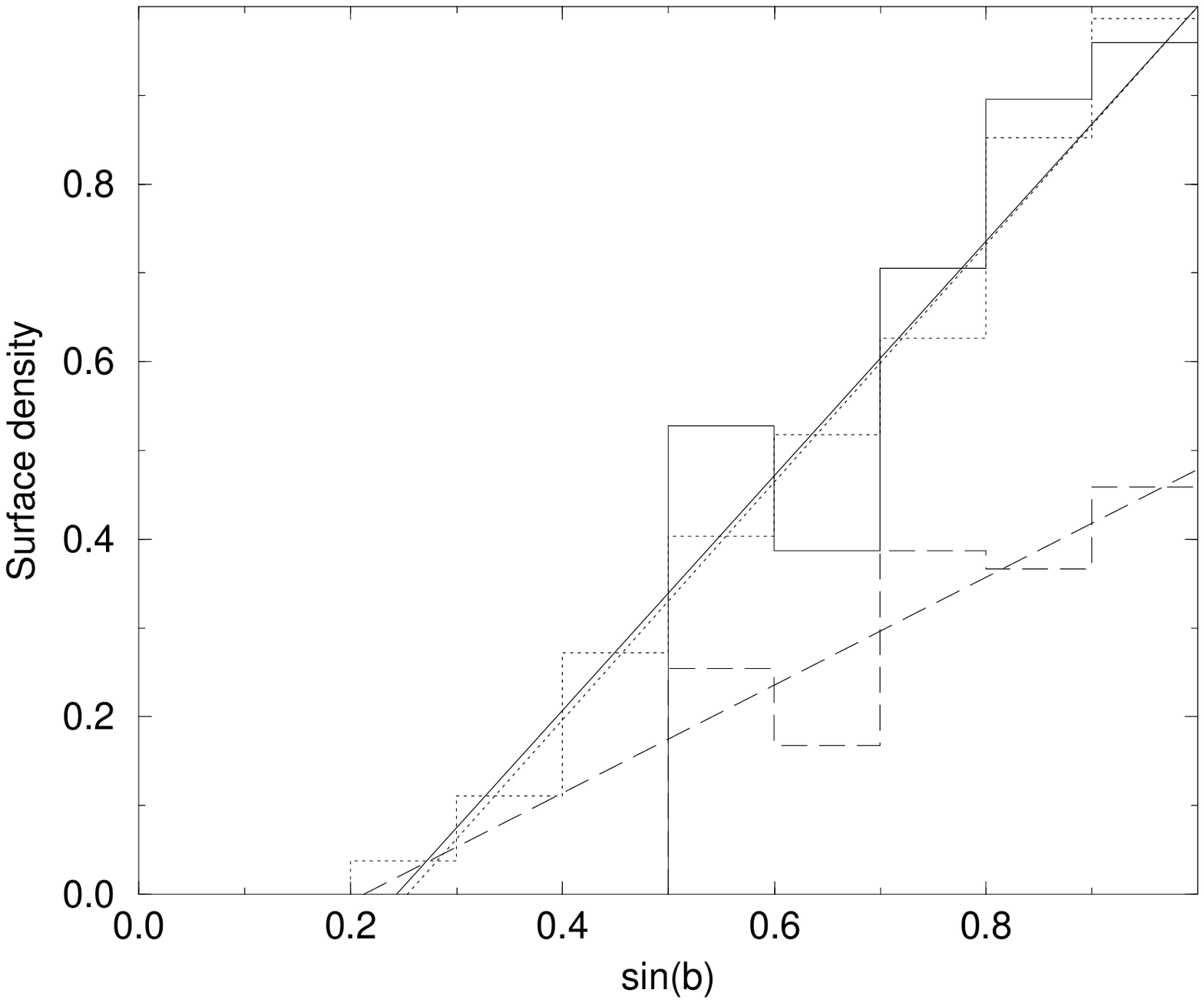}
\label{figure3}
\end{figure*}

In Figure~2 (right-hand panel) we show the number of systems of
various richness in the RBS.250 sample for a neighborhood radius of
$24$~\Mpc.  In total there are 137 clusters in this sample.  At this
neighborhood radius about half of the X-ray clusters are still
isolated, and there are only eight systems with four or more members.
All these systems are embedded in superclusters previously determined
from Abell clusters (E97d, Paper I).  We denote the sample of
superclusters found on the basis of RBS clusters as RBS.XSC.

Actually, a number of X-ray clusters that appear to be isolated in
Figure~2 are members of superclusters of Abell clusters. Thus it is
necessary to determine X-ray clusters in superclusters also in another
way: first we define superclusters on the basis of Abell clusters, and
then find which superclusters contain X-ray clusters (for details and
the lists of superclusters based on Abell X-ray clusters, and
additional superclusters based on non-Abell X-ray clusters see Paper
I).  The sample of members of very rich superclusters among X-ray
clusters obtained in this way is denoted as RBS.A8. The samples
RBS.XSC and RBS.A8 are different. For example, if there is just one
X-ray cluster among the members of a rich supercluster of Abell
clusters then this cluster is included into sample RBS.A8, but not
into sample RBS.XSC.  On the other hand, if in a supercluster with
four members all the members clusters are X-ray clusters then these
clusters are included into sample RBS.XSC, but due to the richness
limit they are not included into sample RBS.A8. In the following we
show the correlation functions for both these samples, RBS.XSC and
RBS.A8.

\section{Correlation function}

\subsection{Selection functions}

Both the Abell and the X-ray cluster sample have a shape of a
double-cone (with the observer at the tip of each cones). The X-ray
cluster sample is bounded by Galactic latitude $|b_g|>30^\circ$ and
extends up to the distance of 250~\Mpc.  Even in this restricted
volume the sample is not homogeneous in spatial density. To take into
account selection effects in the X-ray samples we have determined
galactic latitude and redshift selection functions.  The procedure of
finding of selection functions has been presented in detail in E97b,
E97d, E99a.

Figure~3 shows selection functions for the X-ray (RBS) and Abell
cluster based samples.  The latitude selection can be represented by a 
linear dependence between surface number density $S$ and $\sin b$
\begin{equation}
S(b)=(\sin b - \sin b_0)/(1-\sin b_0),
\label{latit}
\end{equation}
where $b_{0}$ is a parameter in the linear regression.  The distance
dependence can be expressed as a power law 
\begin{equation}
S(d)=\cases{a_0, & $d \le d_0$;\cr\cr
a_0 (d_0/d)^p, & $d > d_0$;}
\label{dist}
\end{equation} 
here $a_0$ is the density of the sample near the observer, $d_{0}$ is
the distance up to which redshift selection is considered constant,
and $p$ is the power law index. Parameters of the selection function for
X-ray and Abell cluster samples are given in Table~2, where $N$ is the
number of clusters in a sample. The distance dependence shows that at
the limiting distance of the sample RBS.250 (250~\Mpc) the density of
the X-ray clusters is about 2.5 times lower than near the observer.
These selection functions were taken into account in the calculation
of the correlation functions for respective samples.

\begin{figure*}[ht]
\vspace*{14.0cm} \figcaption{The correlation functions for X-ray
selected clusters (left-hand panels) and optically selected Abell
clusters (right-hand panels).  Upper panels show correlation functions
for the whole cluster samples RGB.250 and ACO.A1, the lower panels
show clusters which belong to high density regions or superclusters
(samples RBS.A8, RBS.XSC and ACO.A8).  3$\sigma$ error corridors are
shown with dashed lines.  In the lower left panel the solid line is
for X-ray clusters contained in X-ray superclusters (i.e. based on
X-ray clusters only, $N_{Xcl} \geq 2$, sample RBS.XSC); the dashed
line is based on the X-ray clusters in the rich Abell-cluster-based
superclusters ($N_{Acl} \geq $ 8, the sample RBS.A8). }
\includegraphics{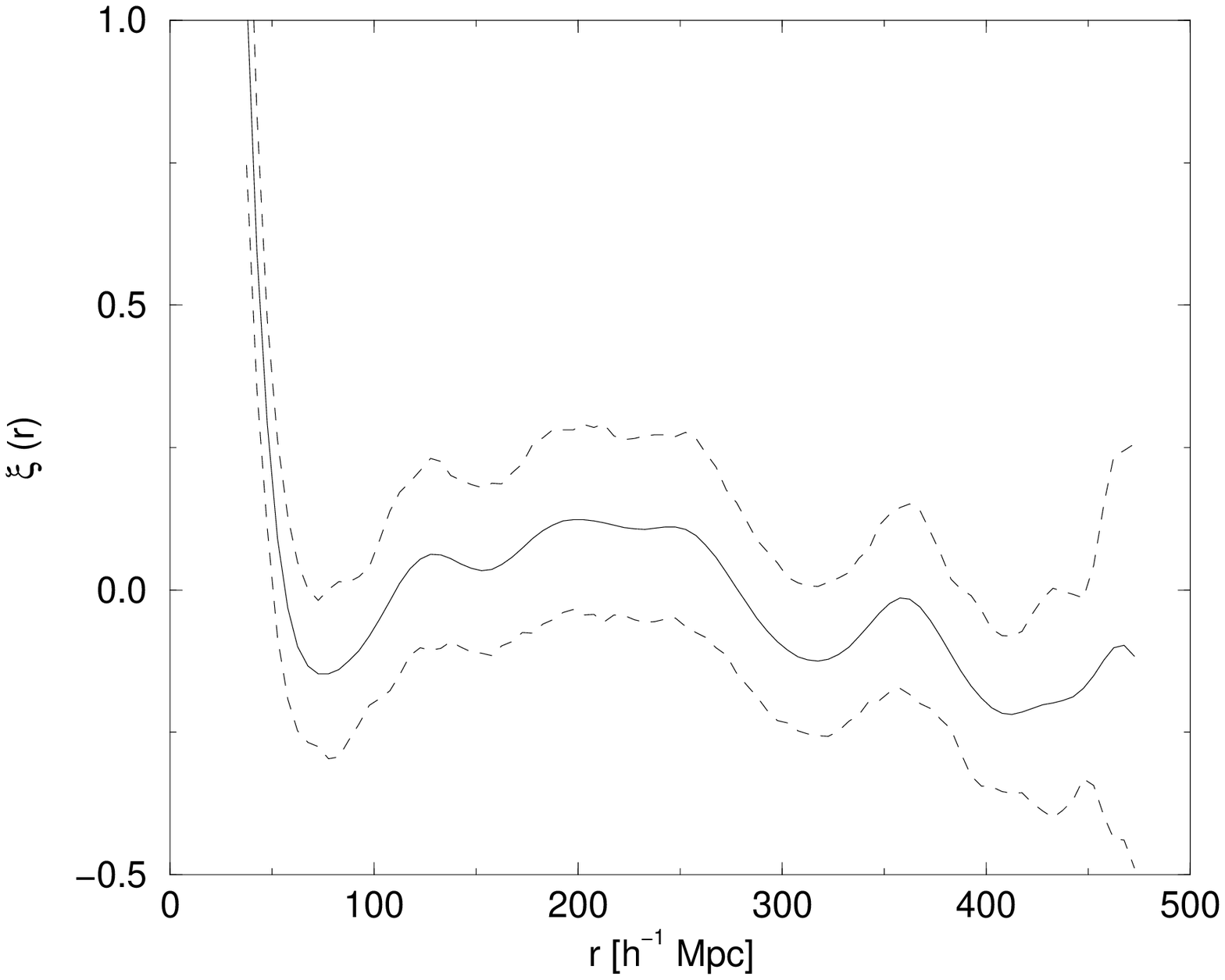}
\includegraphics{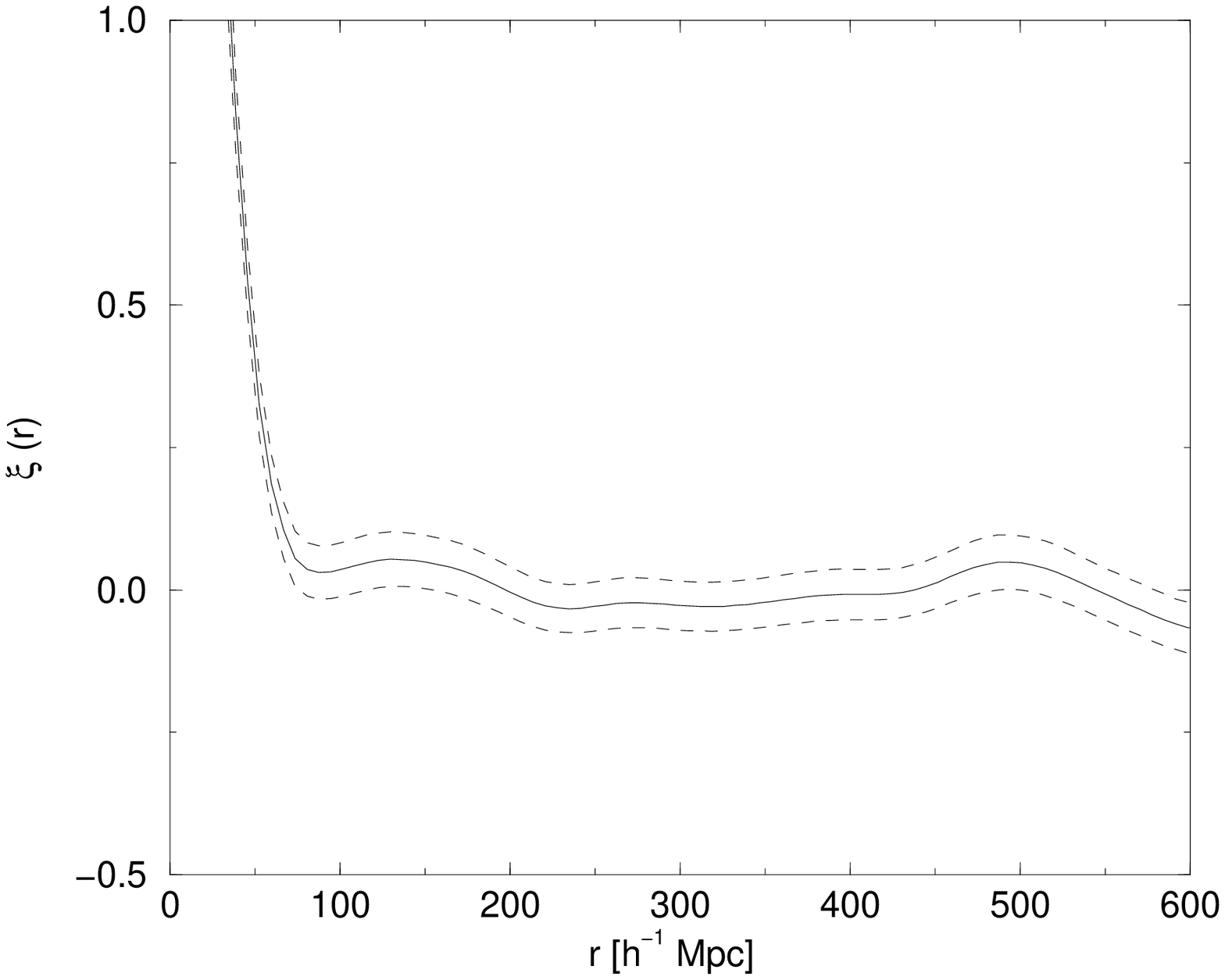}
\includegraphics{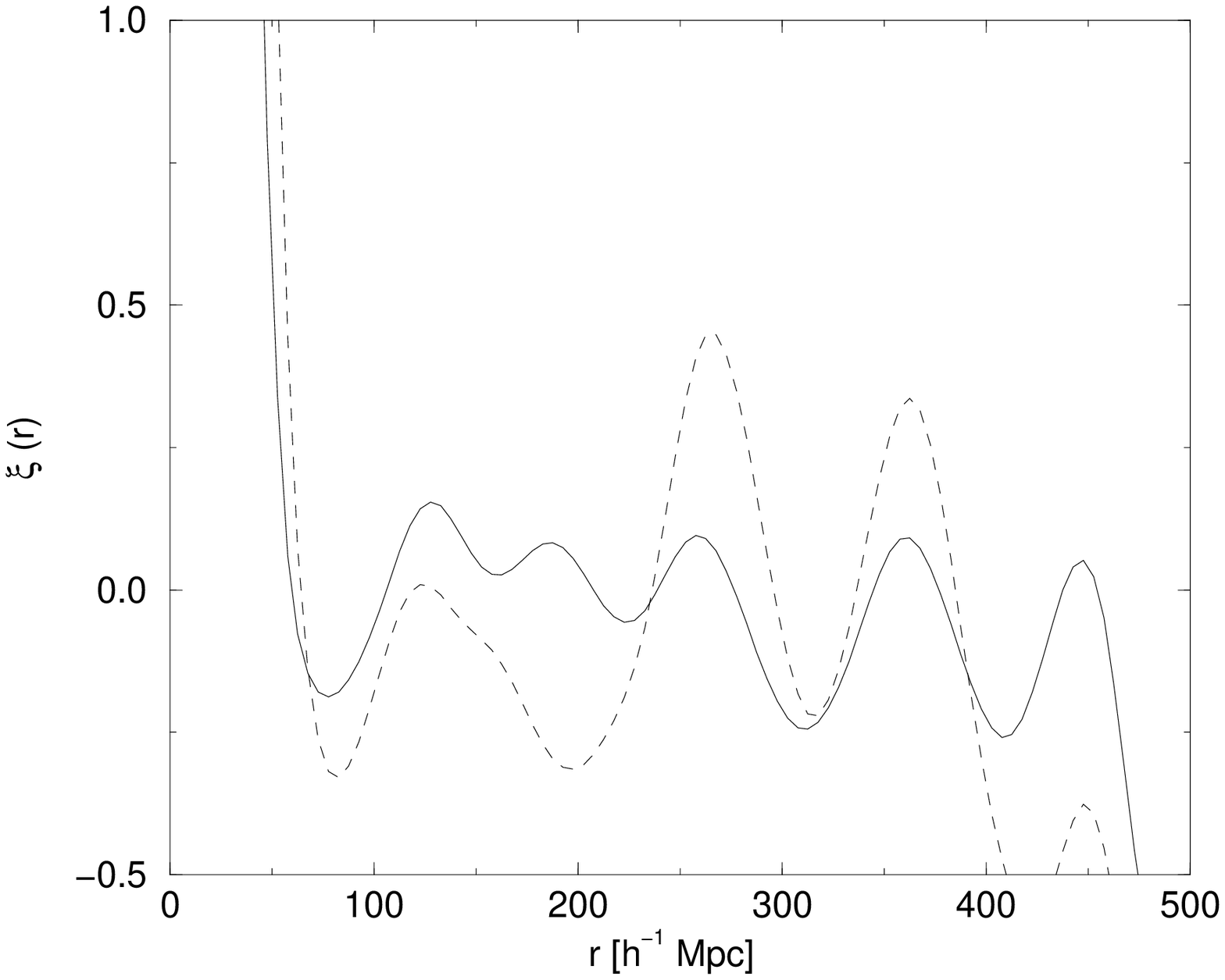}
\includegraphics{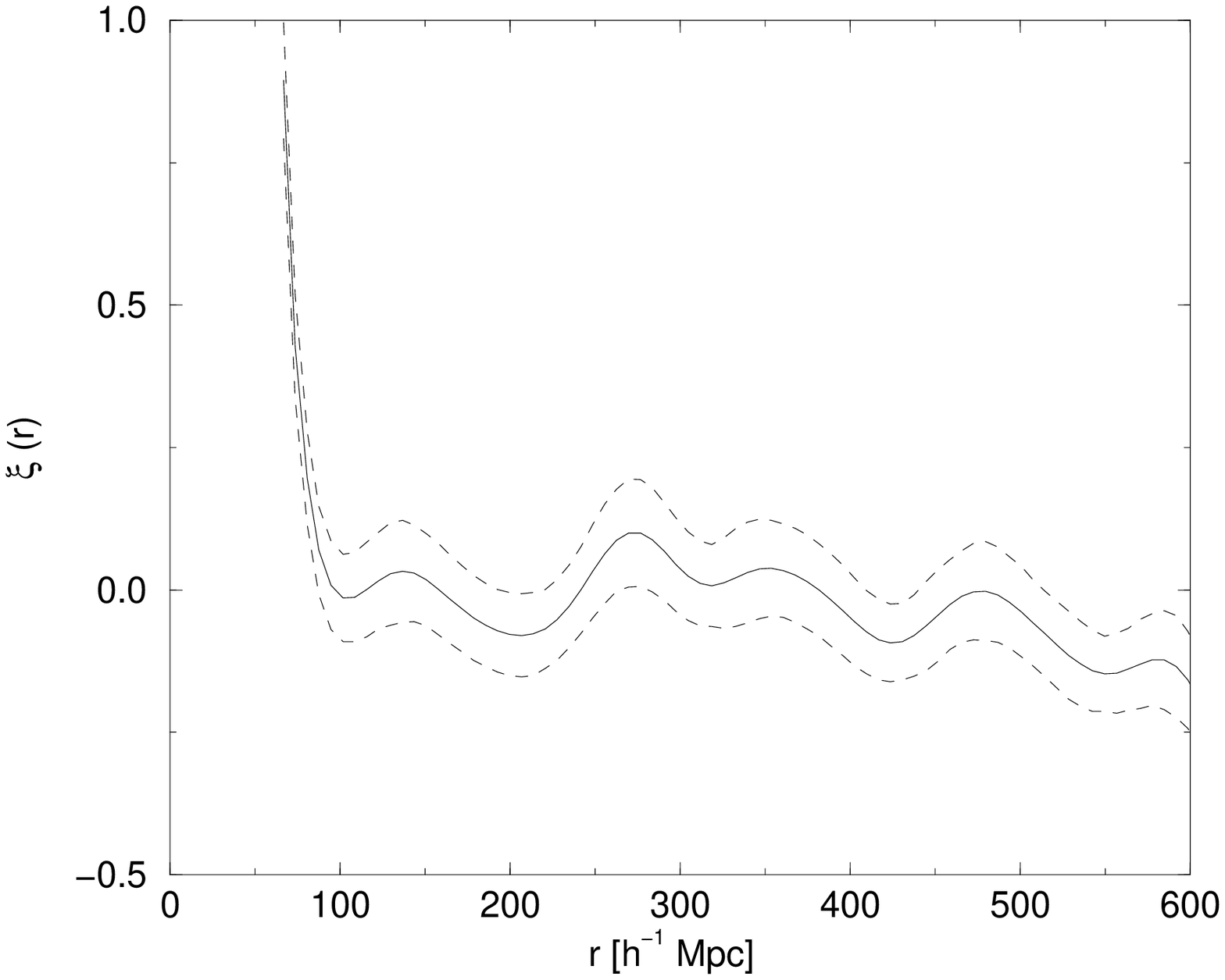}
\label{figure4}
\end{figure*}

\subsection{Correlation functions on large scales}

Now we shall study the correlation function up to scales of 500~\Mpc\
with the purpose to test the presence of oscillations found in
correlation function of optically selected cluster samples.  The
problems which arise in connection with sample volume shape, selection
functions, error estimates and sample dilution have been considered in
detail in our earlier papers (E97b, E97d, E99a).  The two-point
spatial correlation function was calculated in the classical way using
cluster pair separations :
\begin{equation}
\xi(r) = {DD(r) \over RR(r)} - 1,
\label{corr}
\end{equation}
where $DD(r)$ is the number of pairs of clusters in the range of
separations $r\pm dr/2$, $dr$ is the bin size, $RR(r)$ is the
respective number of pairs in a random sample scaled to the mean
number density of clusters in the observed sample. It is assumed that
both samples have identical shape, volume and selection functions. The
mean error was calculated as given in E97b:
\begin{equation}
\sigma_{\xi} = {b \over \sqrt{N}},
\label{err}
\end{equation}
where the value of the parameter $b$ is $\approx$\, 1.5.

\begin{figure*}[ht]
\vspace*{7.0cm} \figcaption{Left panel: correlation functions of Abell
and X-ray selected clusters on a log-log scale; $\xi(r)+1$ is given to
present the correlation function at values $\xi \leq 0$. Right panel:
histogram of distances between centers of rich superclusters of X-ray
clusters.}
\includegraphics{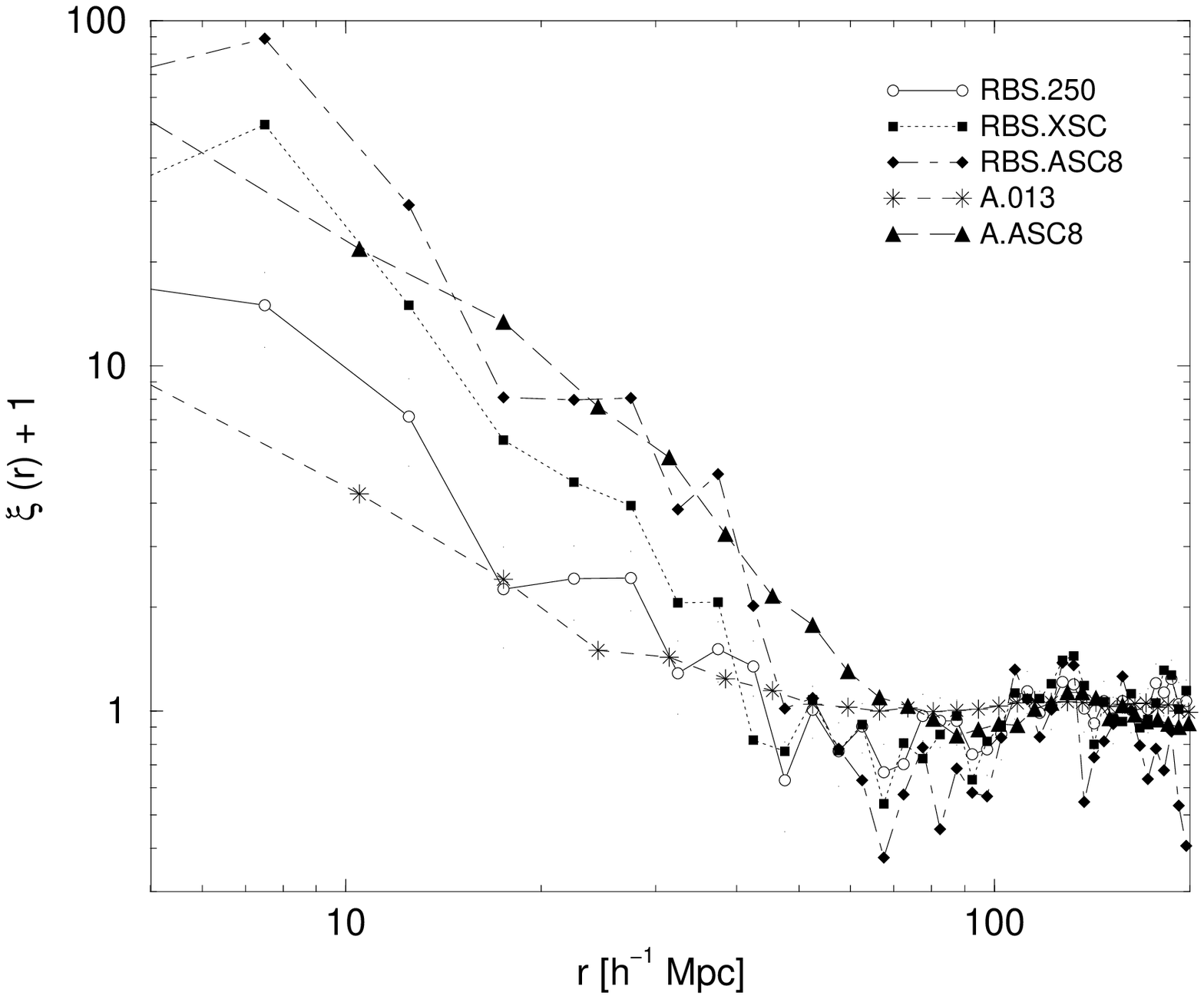}
\includegraphics{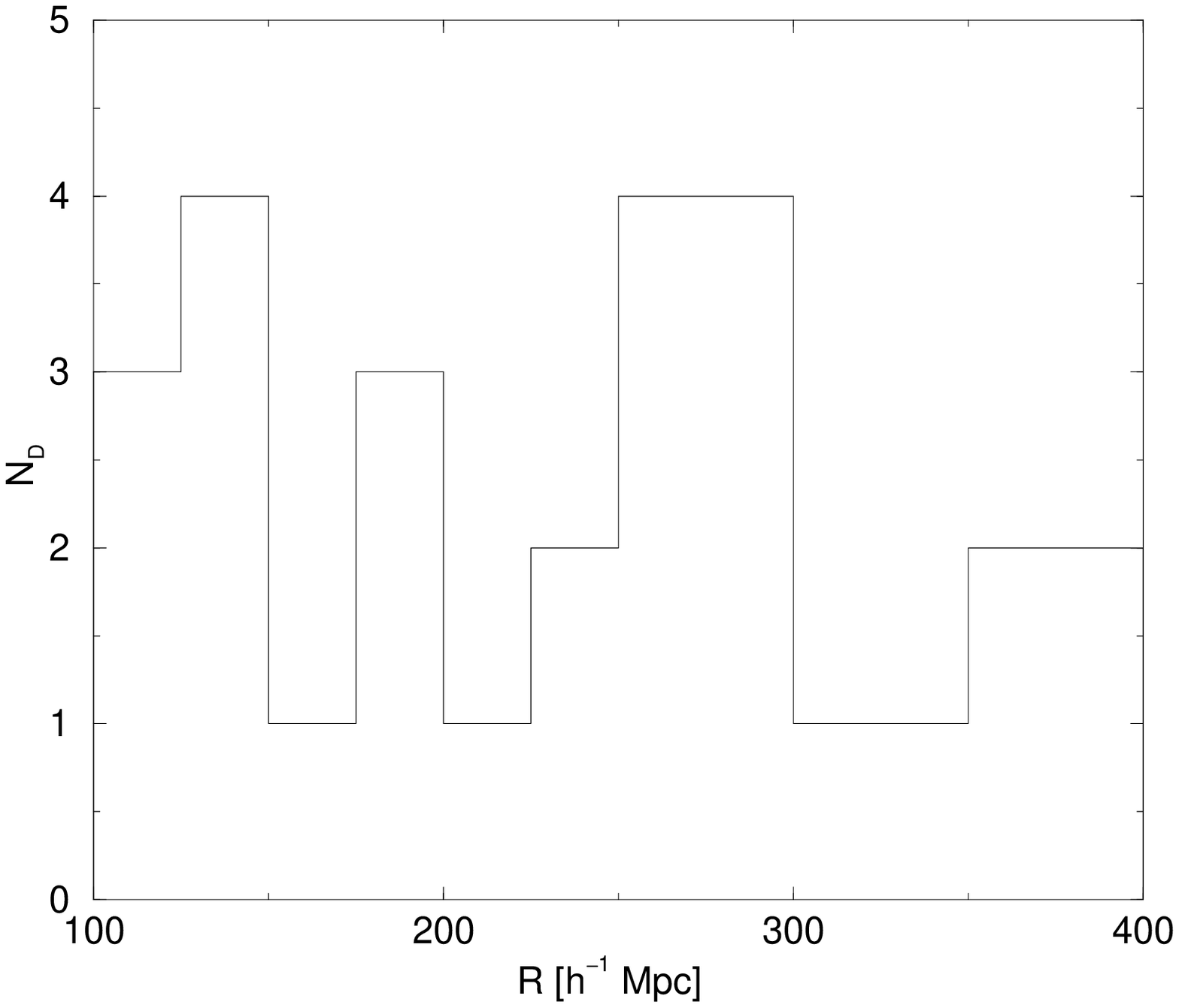}
\label{figure5}
\end{figure*}

In Figure~4 we present the correlation function for the X-ray selected
sample of clusters RBS.250 in comparison with correlation function of
the optically selected (Abell) clusters sample ACO.A1. To decrease
random sampling errors we applied Gaussian smoothing with a dispersion
of 15~\Mpc. Due to the smaller sample size of X-ray clusters the
errors are larger than for the optically selected cluster sample (see
E97d for a more detailed analysis of errors of a correlation
function). The correlation function for X-ray selected clusters shows
the oscillation with a period of about 115~\Mpc\ as in the case of
Abell clusters.  Our earlier studies of optically selected clusters
have shown (E97b) that correlation function of clusters belonging to
high-density regions (i.e. superclusters) is oscillating with a higher
amplitude than the correlation function of all clusters.  In the lower
panels of Figure~4 we present the correlation function of X-ray and
Abell clusters which belong to superclusters as described in section
2.3 (samples RBS.A8, RBS.XSC and ACO.A8).  We see that the amplitude
of oscillations is really higher than for samples of all clusters.  In
the left-hand panels we see five secondary maxima of the correlation
function, in the right-hand panels (which cover a larger range of
separations) even six.  For the X-ray clusters which belong to rich
superclusters the oscillations are more significant than for the
sample of all X-ray clusters.  Mean oscillation periods of the
correlation function for each of our samples are presented in Table 2
(for details on the calculation of periods see E97b).

\subsection{Correlation functions on small scales}

The spatial correlation function of clusters of galaxies on small
scales is usually expressed as a power law
\begin{equation}
\xi = (r/r_0)^{-\gamma},
\label{corr2}
\end{equation}
were $r_0$ is the correlation length, and $\gamma$ the power index.
In the left panel of Figure~5 we present the correlation function for
all observed samples discussed in the present paper; in Table~2 we
give correlation function parameters.  We shall discuss our results in
the last Section.

\section{X-ray clusters in superclusters and the 120 $h^{-1}$ Mpc scale}

In previous Section we saw that the correlation functions for the X-ray
cluster samples belonging to superclusters are oscillating at large
separations with a period of about 115~\Mpc.  E97b and E97d demonstrated
that oscillations of the correlation function at scales larger than
100~\Mpc\ are caused by correlations between clusters in superclusters
at opposite void walls, and appear only if superclusters
form a quasi-regular network of step size which corresponds to the
period of oscillations of the correlation function.

In order to understand better the behavior of the correlation function
of X-ray clusters we shall study the distribution of superclusters of
X-ray clusters.  EETDA and E97d used various independent methods (void
analysis, pencil-beam analysis, the nearest neighbor test) in
addition to the correlation analysis to investigate the geometry of
the supercluster-void network of Abell clusters.  Due to the small
number of X-ray clusters and superclusters we shall restrict our
analysis to study of the distances between supercluster centers for
rich superclusters of X-ray clusters.  The distribution of these
distances is given in the right-hand panel of Figure~5.  We see
several peaks in this distribution, at the distances of about 120,
190, 260 and 370~\Mpc.  We also calculated for each rich supercluster
of X-ray clusters the distance to the nearest neighbor (distances
between supercluster centers). This calculation shows that all rich
superclusters of X-ray clusters have a neighboring rich supercluster
at a distance interval of $100 - 135 $~\Mpc.  Thus peaks in Figure~5
correspond to the distances of the nearest system -- step of the
regularity, diagonal of the regularity and to two steps of the
regularity.  We see that direct calculations of distances between rich
systems of X-ray clusters confirm the presence of the characteristic
scale of $120$~\Mpc\ in the distribution of these clusters seen also
in the correlation function.

We emphasize that the superclusters of X-ray clusters are all embedded
in the superclusters of Abell clusters (see also Paper I), and both
type of superclusters follow the same pattern of large-scale
high-density regions in the Universe.  The small number of X-ray
clusters decreases the significance of this result if we consider
X-ray clusters alone, but in combination with Abell cluster sample we
can find main geometrical properties of the supercluster-void network.

\begin{figure*}[ht]
\vspace*{12.0cm} 
\figcaption{Distribution of X-ray and Abell clusters
in supergalactic coordinates. Filled squares denote X-ray clusters in
rich superclusters, open squares indicate isolated X-ray clusters and
members of poor systems, while open circles mark Abell clusters.  In
order to minimize the projection effects we plot only member clusters
of very rich superclusters with at least 8 members.  The figure shows
all clusters in the range -300~\Mpc\ $\leq$ SGX $\leq$ +300~\Mpc.  }
\includegraphics{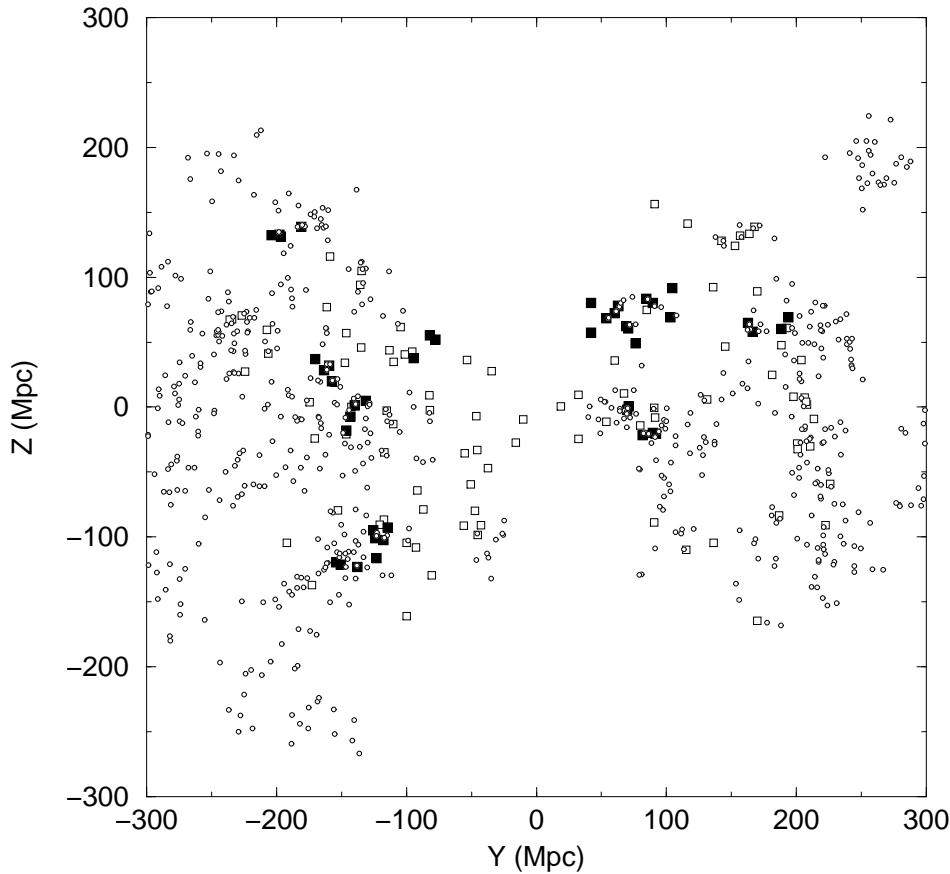}
\label{figure6}
\end{figure*}

In Figure~6 we plot the distribution of X-ray clusters in
supergalactic coordinates; Abell clusters belonging to very rich
superclusters are also shown. Filled symbols correspond to X-ray
clusters in rich systems. We see that the distribution of X-ray
clusters follows the supercluster-void network determined by Abell
clusters (EETDA, E97d). We see a rather regular placement of rich
systems that mainly determine the properties of the correlation
function of X-ray clusters on large scales.

\section{Discussion and conclusions}

It follows from the definition that the correlation function
characterizes the distribution of pairs of clusters at different
separations.  Clusters form superclusters, thus we can say that on
small scales the correlation functions describes the distribution of
clusters within superclusters, and on large scales the distribution of
superclusters themselves.  This property of the correlation function
was discussed by Einasto \etal (1997c). Now we shall discuss in more
detail how properties of superclusters and their distribution manifests
itself in the correlations functions derived in the present paper.

First of all, we notice that different authors use slightly different
algorithms to calculate the correlation function. Also the influence
of various selection effects is treated differently.  In Paper III we
made an independent analysis of the correlation function of Abell
clusters using a slightly different algorithm for the determination of
the selection function. Thus it is appropriate to ask: Are our results
compatible with results of previous investigators?  So far the
correlation function of clusters has been studied mostly up to a scale
of $\sim 100$~\Mpc, and was characterized by parameters of the power
law, the correlation length, $r_0$, and the slope of the correlation
function, $\gamma$.  Values found in the present paper and given in
Table~2 are in the range found previously (see, for example, Bahcall
\& West (1992), Croft \etal (1997), Guzzo \etal (1999), Lee \& Park
(1999), Moscardini \etal (2000), Collins \etal (2000)).  Within the
errors the correlation function parameters for Abell clusters coincide
with values found in Paper III.  The shape of the correlation function
shown in Figure~5 (left-hand panel) is also similar to the shape found
previously by authors noted above.  Thus we can say that our results
are in qualitative agreement with previous studies; in other words,
possible disturbing effects due to differences of the selection
function and algorithms are not dominating our results.

Next we shall discuss differences of parameters of the correlation
function found for different subsamples of Abell and X-ray selected
clusters.  The comparison of correlation functions presented in
Figure~5 (left-hand panel) and parameter values given in Table~2 shows
that the correlation functions of the Abell cluster full sample ACO.A1
and X-ray selected sample RBS.250 are rather similar. The X-ray sample
has slightly higher correlation amplitude and steeper slope, but
within random errors both values are within the range found earlier
for different optically selected cluster samples.  The correlation
function of X-ray clusters in superclusters (sample RBS.XSC) has a
higher amplitude and a larger correlation length.  The highest
amplitude and the largest correlation length are obtained for samples
of clusters in very rich superclusters, RBS.A8 and ACO.A8. Among these
two the X-ray selected sample has a higher amplitude.

These results can be understood as the effect of different threshold
density used in the definition of clusters in superclusters of
different richness, or, in other words, due to different bias levels
of samples.  This aspect has been studied in detail by Einasto \etal
(1999b).  Galaxy and cluster samples are defined by the density field
of matter using various threshold densities to exclude objects located
in a low-density environment.  In galaxy samples the matter in voids
has been excluded, in cluster samples not only the matter in voids but
also the matter in low-density filaments is excluded from the
analysis.  Clusters in superclusters and very rich superclusters
correspond to an increasingly higher density cut-off.  As demonstrated
by Einasto \etal (1999b), a higher density cut-off rises the amplitude
of density fluctuations of objects selected through such a procedure.
This biasing phenomenon is well known since the pioneering work by
Kaiser (1984).

Now we consider the correlation function on large scales. As noted
above, on large scales the correlation function describes the
distribution of superclusters.  If superclusters are randomly
distributed, then the correlation function is approximately equal to
zero for large separations.  On the other hand, if superclusters form
a regular lattice, then the correlation function oscillates with a
period equal to the step size of the lattice (see Einasto \etal 1997c
for a detailed analysis).  We notice that the correlation functions of
all cluster samples have a negative section around $r\approx 80$~\Mpc\
and a secondary maximum on scales of $r\approx 130$~\Mpc.  This
negative section followed by a secondary maximum is a clear indication
for the presence of a dominating scale in the distribution of
superclusters, even if there is no large-scale regularity in the
distribution of superclusters; an example is the Voronoi tessellation
model (Einasto \etal 1997c).  Our cluster samples, both optically and
X-ray selected, have not only this property but show a clear signature
of oscillations with a period $\sim 120$~\Mpc.  The amplitude of
oscillations of X-ray selected clusters is higher than the amplitude
for optically selected Abell clusters in superclusters of similar
richness.  This property shows that cluster samples have not only a
scale but also a regularity in their distribution.  This feature is
stronger in X-ray selected cluster samples, in other words, X-ray
clusters are even better indicators for the location of high-density
regions in the Universe. It is interesting to note that in Paper I we
reached the same conclusion from an analysis of supercluster
memberships: the fraction of isolated X-ray clusters is smaller than
the fraction of isolated Abell clusters, and the fraction of X-ray
clusters in rich superclusters is higher than in poor superclusters.
This is also an evidence that X-ray clusters are better tracers of the
supercluster-void network than optical clusters.  Some other studies
have also shown the presence of a scale of $\sim 120$~\Mpc\ in the
distribution of high-density regions in the Universe (Broadhurst \etal
1990, E97b, E97d, Broadhurst \& Jaffe 1999, Guzzo 1999).

The present study gives no answer to the question: Is the presence of
a scale in the distribution of rich superclusters and their regular
displacement a local or a global phenomenon in the Universe?  
Larger samples of X-ray clusters which shall be available in the near
future shall reveal the distribution of X-ray clusters in a larger
volume and shall yield a better information on the distribution of
high-density regions in the Universe.

To summarize our discussion we can say the following.

1) The correlation analysis has confirmed that both the optically
   selected Abell sample and the X-ray selected RBS sample trace
   high-density regions in the Universe in a similar fashion.

2) Correlation properties of cluster samples depend on the environment
   of clusters: clusters in rich superclusters (high local density)
   have a larger correlation length and amplitude; there exists no
   unique value of the correlation lenth of clusters of galaxies.

3) On large scales the correlation function of clusters depends on the
   distribution of superclusters.  Both optically and X-ray selected
   cluster samples have an oscillating correlation function with
   alternating maxima and minima.  The period of oscillations is $\sim
   115$~\Mpc.  The amplitude of oscillations is larger in
   superclusters of higher richness; for superclusters of similar
   richness the amplitude of oscillations in X-ray selected cluster
   samples is higher than in samples of optically selected clusters.

4) The effective volume covered by the RBS sample is smaller than the
   volume covered by the Abell sample, and the number of known
   clusters is smaller.  The RBS sample is the largest (in volume)
   X-ray selected sample available.  Quantitative statistics based on
   X-ray selected clusters have still a lower significance than those
   based on optically selected samples.  Due to small volume X-ray
   selected samples cannot be considered as fair samples of the
   Universe.
   
5) In combination, optical and X-ray selected samples enhance results
   obtained separately.

\acknowledgments We thank G\"unther Hasinger for providing us with a
draft version of the RBS catalog and discussion of preliminary results
of the study, and Jaak Jaaniste, Enn Saar, Jaan Pelt and Alexei
Starobinsky for stimulating discussion.  This work was supported by
Estonian Science Foundation grant 2625.  JE thanks Astrophysical
Institute Potsdam for hospitality where part of this study was
performed.  HA thanks CONACyT for financial support under grant 27602-E.

\vfill\eject

\vfill\eject
\end{document}